\documentstyle[11pt,newpasp,twoside,psfig]{article}
\begin{document}

\title{New {\it Chandra} Results on Seyfert I galaxies: Fe-K lines}
\author{Urmila Padmanabhan $^{1}$ \& Tahir Yaqoob $^{1,2}$}
\affil{$^{1}$ Johns Hopkins University, 3400 N. Charles St., Baltimore, MD21218.
 \\
$^{2}$ NASA/GSFC, Code 662, Greenbelt Rd., Greenbelt, MD20771.}

\setcounter{page}{1}
\index{Padmanabhan, U.}
\index{Yaqoob, T.}

\begin{abstract}
We present measurements of the Fe-K$\alpha$ line for nine Seyfert I galaxies 
using {\it Chandra} High Energy Transmission Grating Spectrometer ({\it HETGS}) 
data. 
The centroid energies are narrowly dispersed ($6.403\pm0.062$ keV) and  
indicate an origin of the line
cores in cold matter. If {\it all} the lines in this sample 
were the peaks of a 
relativistically broadened disk line, it would require unrealistically fine tuning. 
However, at least three of the nine AGN clearly show a complex Fe-K$\alpha$ line
with an underlying broad component, possibly from a disk. In the
others, an apparently narrow 
Fe-K$\alpha$ line (even if it is resolved by {\it Chandra}) may still be due to 
the peak of a disk line. To distinguish this scenario from an origin in distant 
matter requires variability information.
\end{abstract}

\vspace{-0.8cm}
\section{Introduction}
The Fe-K$\alpha$ emission lines in type 1 AGN can be complex, consisting of a 
broad ($\sim 10,000-100,000$ km/s FWHM) and/or a narrow ($< 10,000$ km/s FWHM) 
component (e.g. Yaqoob et al. 2002). The broad component probably originates 
from the accretion disk, having suffered severe Doppler and gravitational energy
 shifts in the inner disk (e.g. see review by Fabian et al. 2000) and has been 
clearly observed by {\it ASCA}. {\it Chandra} and {\it XMM} are revealing that a
 narrow-line (NL) component is quite common in type 1 AGN (Reeves et al. 2001; 
Yaqoob et al. 2001a; Yaqoob et al. 2001b; Kaspi et al. 2001; Pounds et al. 2001;
 Turner et al. 2002, Fang et al. 2002), often superimposed on top of a broad 
component. Even with {\it Chandra}, the NL component has only been resolved in 
two cases at $> 99\%$ confidence (NGC 3783 [Kaspi et al. 2002] and MCG~$-$6$-$30$-$15 [Lee et al. 2002]). Interpreting the line width in NGC 3783 ($1720\pm360$ 
km/s FWHM) as the result of Doppler shifts in a spherical gas distribution with
Keplerian velocity places its origin in between the BLR and NLR. However, variability 
information is required to truly constrain the origin of the line. The NL 
observed in MCG~$-$6$-$30$-$15 has a width of $\sim$ 3600 km/s FWHM (high flux 
state) but it cannot arise from distant matter, as demonstrated by the rapid 
variability of the line (Lee et al. 2002). Thus, some of the narrow lines 
observed by {\it Chandra} could be the cores of an underlying broad disk line.  

Here we examine the data from nine Seyfert I galaxies observed by the 
{\it Chandra HETGS}. We present the most precise measurements to date of the 
peak energies and widths of the line cores. By comparing the confidence contours
 of the equivalent widths (EW) versus FWHM velocity widths and line intensity 
versus peak energy, we can infer important differences about the origin of the 
lines in different sources.

\section{Method}
The analysis method was similar to that used in Yaqoob et al. (2001a) except that
 the data were reprocessed using {\tt ciao 2.1.3} and {\tt CALDB} version 2.7, 
according to recipes described in {\tt ciao 2.1.3}. The {\it HETGS} is composed of the
 {\it High Energy Grating} (HEG) and the {\it Medium Energy Grating} (MEG) and 
we used only the HEG data since it has the best energy resolution (1860 km/s
FWHM at 6.4 keV) in the Fe-K$\alpha$ region. We fitted the HEG data with a model
 of simple power law 
plus a Gaussian from 2 keV to 7 keV. All parameters, including the line width, 
were allowed to float. The power law slope was then frozen at the best-fit 
value. Using just the 5 to 7 keV data, the joint confidence contour plots of the
 line intensity versus the line energy, and EW versus the FWHM were made. These 
are shown in Figures 1 and 2 respectively. The best-fitting line parameters, with
 statistical errors at $90\%$ confidence for one interesting parameter, are shown
 in Table 1. Note that the results obtained when the power-law slope is allowed 
to float are similar (full details will be presented elsewhere). 

\vspace{-0.4cm}
\begin{table}

\caption{Power law plus Gaussian fits to the $\it Chandra$ (HEG) data.}
\begin{tabular}{lcccc}
\tableline
Source & $\rm E^{a}$ & $ \rm I^{b}$ & $ \rm EW^{c}$ & FWHM  \\
 & (keV) & ($\rm 10^{-5} \ photons \ cm^{-2} \ s^{-1}$) & (eV) & (km/s) \\
\tableline

NGC 3516 & $6.400^{+0.008}_{-0.010}$ & $4.3^{+1.3}_{-1.1}$ & $162^{+47}_{-41}$ & $1390^{+1640}_{-1390}$ \\

NGC 3783 & $6.399_{-0.013}^{+0.013}$  & $5.3_{-1.8}^{+2.2}$ & $79^{+33}_{-27}$ &$2640_{-1265}^{+2200}$ \\

NGC 5548 & $6.397^{+0.021}_{-0.022}$ & $3.0^{+1.5}_{-1.3}$ & $106^{+53}_{-46}$ &$3650^{+2690}_{-2000}$  \\

3C 120 & $6.414^{+0.019}_{-0.016}$ & $3.0^{+1.8}_{-1.4}$ & $61^{+36}_{-29}$ & $2090^{+2860}_{-1920}$ \\

Mkn 509 & $6.431^{+0.027}_{-0.023}$ & $2.8^{+1.9}_{-1.7}$ & $45^{+30}_{-27}$ & $2740^{+2630}_{-2740}$ \\

NGC 4593 & $6.402^{+0.014}_{-0.020}$ & $3.2^{+1.5}_{-1.3}$ & $73^{+35}_{-29}$ & $2140^{+4030}_{-1480}$ \\

NGC 4051 & $6.417^{+0.035}_{-0.031}$ & $2.7^{+1.6}_{-1.3}$ & $147^{+86}_{-69}$ & $5190^{+5530}_{-3540}$ \\

F 9 & $6.424^{+0.031}_{-0.089}$ & $2.2^{+2.9}_{-1.1}$ & $86^{+113}_{-43}$ & $4040^{+15720}_{-1840}$  \\

3C 273 & $6.348^{+0.052}_{-0.048}$ & $5.5^{+5.9}_{-1.5}$ & $32^{+35}_{-9}$ & $4610^{+8050}_{-4170}$ \\

\tableline
\tableline
\end{tabular}
Statistical errors are for 90\% confidence for one interesting parameter ($\Delta C$ = 2.706). $^{a}$ Center energy of Gaussian line. $^{b}$ Line Intensity. $^{c}$ Equivalent Width.
\end{table}

\vspace{-0.6cm}
\section{Centroid Energies of the Fe-K lines}
In Figure 1 we show the line intensity versus line center energy confidence 
contours. Note
 that the dotted line is at 6.4 keV and all the contours intersect it, at least 
at the $99\%$ confidence level. Due to the superb energy resolution of the 
{\it Chandra} HEG, the $90\%$ statistical errors on the line energy can be as 
small as 10 eV. The most tightly constrained contours are those of NGC 3516, 
which has an exposure time of 100 ks and a comparatively low (2$-$10 keV) flux 
of $2.5 \times 10^{-11} \rm \ erg \ cm^{-2} \ s^{-1}$. The weighted mean centroid 
energy of the line detected in the nine Seyfert galaxies is $6.403\pm0.062$ keV.
 The small dispersion is quite remarkable. If all these lines were from a disk, 
the disk parameters would need to be finely tuned. Whatever the origin of the 
lines, the cores observed by the HEG are formed in cold Fe.

The detailed shapes of the contours carry additional information about the line 
profiles. Comparing the contours of NGC 5548 and F 9, we see a clear and marked 
difference in the shapes of their contours. However, they have a similar 
signal-to-noise ratio (SNR) as evidenced by the number of photons in the 5 to 7 
keV band. Thus, the difference in the contours are due to the intrinsic differences 
in their line profiles and {\it are not due to SNR effects}. The broader 
contours of F 9 indicate a complex line, consistent with the line profile in 
Yaqoob et al. (2001b). {\it RXTE} data, simultaneous with the {\it Chandra} 
observation for F 9, confirms the presence of a broad He-like Fe-K$\alpha$ 
component (see Yaqoob \& Padmanabhan 2003; hereafter YP03). The SNR in the other
 seven AGN is greater than that of NGC 5548 and F 9. Therefore, we can infer 
that the 
differences in the contours of {\it all} the AGN are not due to SNR effects, 
except for 3C 273 where the EW of the line is much less than in the other 
sources. The contours of NGC 3516, NGC 3783, 3C 120 and MKN 509 are very similar
 to that of NGC 5548 and those of NGC 4051 and NGC 4593 are similar to that of 
F 9.  

\vspace{-0.2cm}
\section{EW versus FWHM}
In Figure 2 the EW versus FWHM confidence contours are shown. Note that the 
dotted line on these plots shows the FWHM resolution of the HEG (1860 km/s at 
6.4 keV for $z$ = 0) at the measured, {\it observed} peak energy of each line 
(that is, accounting for redshift). 
Again, comparing NGC 5548 and F 9 (since they have similar SNR), we find that 
their contours have very different shapes. The contours of NGC 5548 show the 
99\% confidence level (CL) FWHM bound at $\sim$ 10,000 km/s with the line peak 
 unresolved at the $99\%$ CL. On the other hand, in F 9 the FWHM extends out 
$\sim$ 60,000 km/s at the $99\%$ CL (which is off the scale in Figure 2 but can
be seen in YP03). F 9 also shows the only resolved peak at
 the $99\%$ CL in this sample. Thus, the lines in F 9 and NGC 5548 are 
very clearly different, with the line in F 9 definitely accompanied by a broader
 component and the line in NGC 5548 comparatively narrower. 
The FWHM contours of NGC 3516, NGC 3783, 3C 120, and Mkn 509 are similar to that
 of NGC 5548. NGC 4593, NGC 4051 and F 9 all show clear underlying broad
components. Except for F 9, all the Fe-K$\alpha$ lines in the sample, show an 
unresolved peak at the 99\% CL.

Lee et al. (2002) found that in a {\it Chandra} observation of MCG~$-$6$-$30$-$15, the
HEG Fe-K$\alpha$ line was resolved, with a best-fit FWHM width of 11,000 km/s 
(time-averaged). Considering only the high flux state, the line was resolved 
with a best-fit FWHM width of 3600 km/s. However, this line was shown to be 
variable on short time scales (hundreds of seconds). Thus, this line is most 
probably the core of a very peaky disk line, since it is impossible for either 
the putative obscuring torus, or even the BLR to respond to continuum variations
 so quickly. This demonstrates that even the AGN in our sample which show 99\% 
CL FWHM contours less than 15,000 km/s may have Fe-K$\alpha$ line cores which 
are from a disk and not from more distant matter.   
  
On the other hand, for the dispersion in the line centroid energies to be so
small, the radial emissivity of the disk must be very flat {\it and} the disk 
inclination angles must be small and narrowly distributed (almost face-on). To 
determine the dominant contribution to the line core (disk or distant matter) 
requires variability information. The EWs of the line cores range from $\sim$
30 eV to $\sim$ 150 eV. When there is an additional broad line, as measured by 
a higher throughput mission (e.g. {\it RXTE}; see YP03), the total EW can easily
exceed twice this value. However, an interpretation of the EWs requires 
detailed physical modeling which will be presented elsewhere.  

\vspace{-0.3cm}
\section{Conclusions}
Measurements of the Fe-K$\alpha$ line centroid 
energy and width for nine Seyfert I galaxies using the best energy resolution
available show that the centroid energies 
of the line cores are narrowly 
dispersed ($6.403\pm0.062$ keV) and are consistent with a dominant contribution
{\it to the line core} from cold matter. Three sources (NGC 4051, NGC 4593 and F 9) 
clearly show a complex Fe-K$\alpha$ line with an underlying broad, relativistic 
component. In the remaining AGN (except 3C 273), although the FWHM is
$ <$ 11,000 km/s (99\% CL), we cannot rule out the line core still coming from a
 disk (as opposed to originating in distant matter). However, the small 
dispersion in the peak energy requires a very flat radial emissivity law and a
 narrowly distributed disk inclination angle. Variability information is 
required to determine the origin of the line core in these cases.

\acknowledgements
The authors gratefully acknowledge support from
NASA grants NCC-5447 (T.Y., U.P.), NAG5-10769 (T.Y.),
and CXO grants GO1-2101X, GO1-2102X (T.Y.).
The authors are grateful to the {\it Chandra} and {\it RXTE}
instrument and operations teams for their work.
We also thank our collaborators, B. McKernan, I. M. George,
T. J. Turner, \& K. Weaver.

\begin{figure}[s]
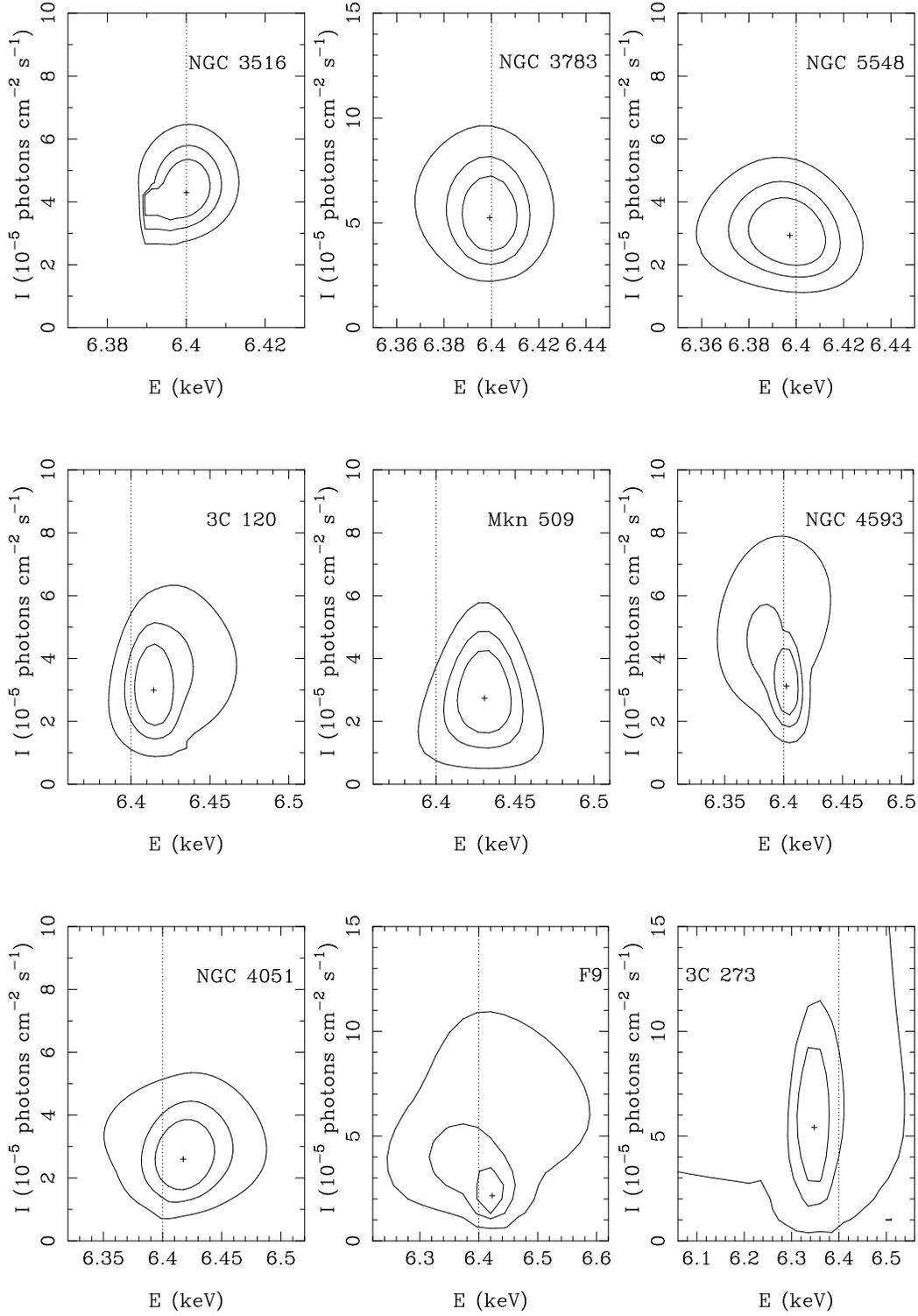

\centerline{
\psfig{figure=urmila_n3516_en_2.ps,width=4.5cm,height=6cm}
\psfig{figure=urmila_n3783_en_2.ps,width=4.5cm,height=6cm}
\psfig{figure=urmila_n5548_en_2.ps,width=4.5cm,height=6cm}
}
\vspace{0.9cm}
\centerline{
\psfig{figure=urmila_3c120_en_2.ps,width=4.5cm,height=6cm}
\psfig{figure=urmila_m509_en_2.ps,width=4.5cm,height=6cm}
\psfig{figure=urmila_n4593_en_2.ps,width=4.5cm,height=6cm}
}
\vspace{0.9cm}
\centerline{
\psfig{figure=urmila_n4051_en_2.ps,width=4.5cm,height=6cm}
\psfig{figure=urmila_f9_en_2.ps,width=4.5cm,height=6cm}
\psfig{figure=urmila_3c273_en_2.ps,width=4.5cm,height=6cm}
}

\caption{HEG confidence contours (68\%, 90\%, 99\%) of Fe-K line of Line Intensity versus Line Energy for the nine AGN. }
\label{fig1}
\end{figure}

\begin{figure}[t]
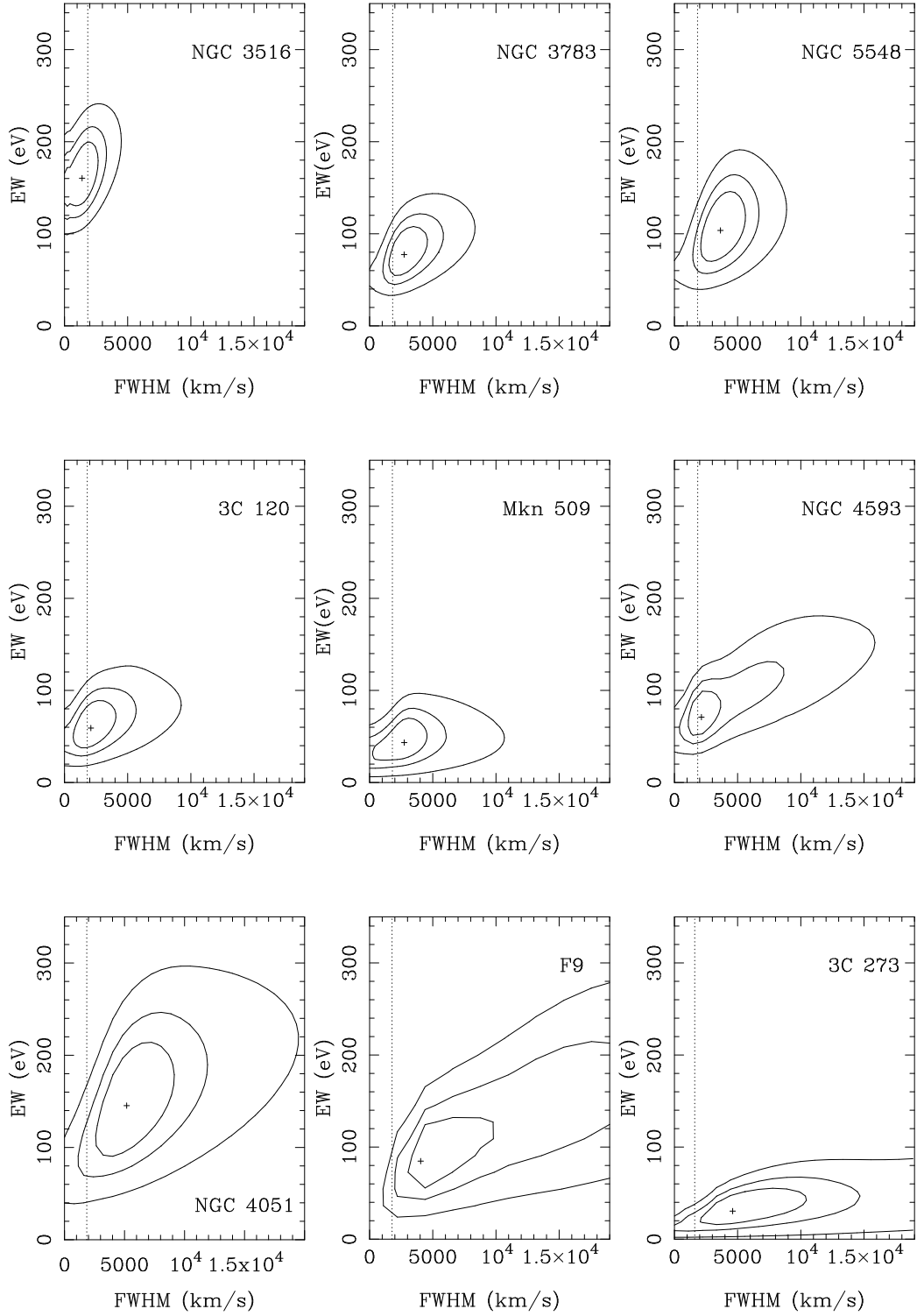

\centerline{
\psfig{figure=urmila_n3516_fwhm_bw.ps,width=4.5cm,height=6cm}
\psfig{figure=urmila_n3783_fwhm_bw.ps,width=4.5cm,height=6cm}
\psfig{figure=urmila_n5548_fwhm_bw.ps,width=4.5cm,height=6cm}
}
\vspace{0.9cm}
\centerline{
\psfig{figure=urmila_3c120_fwhm_bw.ps,width=4.5cm,height=6cm}
\psfig{figure=urmila_m509_fwhm_bw.ps,width=4.5cm,height=6cm}
\psfig{figure=urmila_n4593_fwhm_bw.ps,width=4.5cm,height=6cm}
}
\vspace{0.9cm}
\centerline{
\psfig{figure=urmila_n4051_fwhm_bw.ps,width=4.5cm,height=6cm}
\psfig{figure=urmila_f9_fwhm_bw.ps,width=4.5cm,height=6cm}
\psfig{figure=urmila_3c273_fwhm_bw.ps,width=4.5cm,height=6cm}
}

\caption{HEG confidence contours (68\%, 90\%, 99\%) of Fe-K line  EW vs. FWHM of
the nine AGN. }
\label{fig2}
\end{figure}

\end{document}